\RequirePackage[2020-02-02]{latexrelease}
\documentclass[reprint,
groupedaddress,
unsortedaddress,
 amsmath,amssymb,
 aps,
superscriptaddress
]{revtex4-2}
\usepackage[textsize=tiny]{todonotes}
\usepackage{float}
\usepackage{graphicx}
\usepackage{dcolumn}
\usepackage{bm}
\usepackage{hyperref}
\usepackage[normalem]{ulem}
\usepackage{xr}
\makeatletter
\newcommand*{\addFileDependency}[1]{
  \typeout{(#1)}
  \@addtofilelist{#1}
  \IfFileExists{#1}{}{\typeout{No file #1.}}
}
\makeatother

\newcommand*{\myexternaldocument}[1]{%
    \externaldocument{#1}%
    \addFileDependency{#1.tex}%
    \addFileDependency{#1.aux}%
}
\myexternaldocument{SI}

\makeatletter
\newcommand*{\rom}[1]{\expandafter\@slowromancap\romannumeral #1@}
\makeatother

\def\kt{k_\text{B}T}
\def\EB{E_\text{B}}

\def\B{B}
\def\N{N}

\def\t{\tau}
\def\kg{k_\text{grow}}

\def\kgo{k_\text{grow}^0}
\def\kc{k_\text{close}}
\def\kco{\kc^0}
\def\muEq{\mu_{\text{eq}}}
\newcommand*{\kcNet}{\Tilde{k}_\text{close}}

\def\Pop{P_\text{open}}

\newcommand*{\PClose}{P_\text{close}}

\newcommand*{\kStretch}{k_\text{S}}
\newcommand*{\aMono}{a_0}
\newcommand*{\thetaIdeal}{\theta_{\text{id},i}^{(m,n)}}
\newcommand*{\thetaIdealTarget}{\theta_{\text{id},i}^{0}}
\newcommand*{\LEnd}{L_\text{end}}
\newcommand*{\LClosure}{L_\text{close}}
\newcommand*{\NClosure}{N_\text{close}}

\newcommand*{\fFusion}{f_\text{fusion}}

\newcommand*{\GClose}{G_\text{close}}
\newcommand*{\IClose}{I_\text{close}}

\newcommand*{\NMax}{N_\text{max}}

\newcommand*{\MeanD}{\Bar{D}}

\newcommand*{\rhoFree}{c_0}
\newcommand*{\cSS}{c_\text{SS}}

\begin{document}

\preprint{APS/123-QED}

\title{Polymorphic self-assembly of helical tubules is kinetically controlled}

\author{Huang Fang}
\affiliation{Martin Fisher School of Physics, Brandeis University, Waltham, Massachusetts 02454, USA}
\author{Botond Tyukodi}
\affiliation{Martin Fisher School of Physics, Brandeis University, Waltham, Massachusetts 02454, USA}
\affiliation{Department of Physics, Babe\c{s}-Bolyai University, 400084 Cluj-Napoca, Romania}
\author{W. Benjamin Rogers}
\affiliation{Martin Fisher School of Physics, Brandeis University, Waltham, Massachusetts 02454, USA}
\author{Michael F. Hagan}
\affiliation{Martin Fisher School of Physics, Brandeis University, Waltham, Massachusetts 02454, USA}

\date{\today}

\begin{abstract}
In contrast to most self-assembling synthetic materials, which undergo unbounded growth, many biological self-assembly processes are self-limited. That is, the assembled structures have one or more finite dimensions that are much larger than the size scale of the individual monomers. In many such cases, the finite dimension is selected by a preferred curvature of the monomers, which leads to self-closure of the assembly.  In this article, we study an example class of self-closing assemblies: cylindrical tubules that assemble from triangular monomers. By combining kinetic Monte Carlo simulations, free energy calculations, and simple theoretical models, we show that a range of programmable size scales can be targeted by controlling the intricate balance between the preferred curvature of the monomers and their interaction strengths. However, their assembly is kinetically controlled --- the tubule morphology is essentially fixed shortly after closure, resulting in a distribution of tubule widths that is significantly broader than the equilibrium distribution. We develop a simple kinetic model based on this observation and the underlying free-energy landscape of assembling tubules that quantitatively describes the distributions. Our results are consistent with recent experimental observations of tubule assembly from triangular DNA origami monomers. The modeling framework elucidates design principles for assembling self-limited structures from synthetic components, such as artificial microtubules that have a desired width and chirality.
\end{abstract}


\maketitle

\section{Introduction}

Many biological functions rely upon the assembly of \emph{self-limited} structures that have well-defined finite sizes, and yet are much larger than the size of the individual building blocks. 
Examples include the assembly of protein capsomers into viral shells with the appropriate size to encapsulate the viral nucleic acid, assembly of tubulin into microtubules with diameters that confer sufficient rigidity to mechanically support the cell~\cite{Weisenberg1972,Desai1997}, and, within butterfly wings, the organization of chitin into nanostructured domains on the scale of visible light to make the tissue iridescent~\cite{Han2009,Saranathan2010}. 
In contrast, most structures assembled from synthetic building blocks undergo \emph{unlimited} growth into crystals or amorphous materials~\cite{Manoharan2015,Boles2016,lee2012}. 
The biological structures described above are examples of `curvature-controlled' assemblies, in which the building blocks assemble with a preferred curvature that leads the structure to close upon itself in one or more directions.

There has been an intense interest in mimicking such functional biological structures by developing synthetic building blocks that can be pre-programmed to assemble with curvatures leading to self-closure. To this end, researchers have recently used DNA origami (e.g.~\cite{Sigl2021,Hayakawa2022}) and protein design (e.g.~\cite{Bale2016, Golub2020, Butterfield2017}) to engineer building blocks that assemble into polyhedral capsids or tubules with designed diameters. However, due to thermal fluctuations and kinetic effects, assembled structures typically exhibit polymorphism in the limited dimension rather than a single well-defined diameter ~\cite{lavelle2009,mohammed2013,Hagan2021}. Understanding the factors that control this size distribution is essential for achieving functional self-limited assemblies. In this article, we use computer simulations and kinetic models to understand the dynamical pathways of helical tubule assembly, and the resulting polymorphic distribution of assembled tubule structures.

Curvature-controlled assemblies in biology frequently rely on symmetry principles to maximize their `economy' of assembly, meaning the size of the structure that can be assembled for a given number of distinct subunit species~\cite{Hagan2021}. For example, icosahedral symmetry maximizes the number of identical subunits (60) that can be used to assemble a shell, and many viruses assemble icosahedral capsids~\cite{Caspar1962, Johnson1997, Dokland2000, Natarajan2005}. In this sense, helical tubules are even simpler than icosahedral capsids --- there is an infinite family of helical tubules with different diameters and pitches, each of which can be assembled from a single subunit species with identical conformations throughout the structure. However, because the subunit curvature changes only slightly between different tubule structures with similar geometries within this family, tubule assembly is highly susceptible to polymorphism. That is, when subunits associate with imperfect geometries during assembly, and these imperfections fail to anneal before becoming trapped by further subunit association, the resulting assembled structures deviate from the ground state tubule structure. Consequently, the geometry distribution of tubules assembled in a finite time depends on a competition between kinetic and thermodynamic factors, and can differ significantly from the equilibrium distribution. Identifying these factors from experiments alone is challenging because most intermediate structures are transient and present at concentrations which are too low to experimentally detect or characterize.

Computer simulations can help to understand self-limited size distributions by revealing the dynamical pathways leading to assembly. 
However, in comparison to the extensive body of theoretical and computational modeling of icosahedral capsids or shells (e.g.~\cite{hagan2016}), there has been relatively limited study of tubule assembly (e.g.~\cite{Cheng2012,Cheng2013,Cheng2013a,Bollinger2018,Bollinger2019,Stevens2017,Kraft2012,Marchetti2019,HernandezGarcia2014}). Thus, the mechanisms controlling tubule assembly and  closure have yet to be completely explored. 

In this article, we perform kinetic Monte Carlo simulations on a model of triangular subunits motivated by recent experiments demonstrating the assembly of DNA origami building blocks into helical tubules~\cite{Hayakawa2022}. By comparing the distribution of dynamically assembled tubules with equilibrium results, we find that the size and morphology distribution is kinetically controlled. In particular, the structural ensemble is typically quenched shortly after a nascent assemblage first closes upon itself to form a cylindrical tubule. Through a combination of dynamical simulations, free-energy calculations, and simple analytical models, we determine how the resulting size distribution depends on control parameters such as the bending modulus and the pre-programmed target curvature. 
These results may guide the experimental design of more efficient and accurate self-assembling artificial tubule structures.

The remainder of the article is organized as follows: In section~\ref{sec:Sim}A--B, we introduce the kinetic Monte Carlo simulation that we use to model tubule self-assembly. We then discuss the predicted assembly trajectories and geometry distribution of assembled tubules. In section ~\ref{sec:SimvsExp}, we compare simulation outcomes to observations from experiments on tubules self-assembled from DNA origami subunits, and obtain an estimate of the bending rigidity in the experimental system.  In section ~\ref{Sec:Theory}A--C, we present calculations of the equilibrium tubule geometry distribution and, through comparison with simulation results, show that the assembled geometry distribution is kinetically controlled. In section ~\ref{Sec:Theory}D--E, we construct a kinetic model that captures these kinetic effects, and use it to predict the assembly behavior as a function of the control parameters.
Finally, in section~\ref{sec:conclusions}, we discuss implications for future experiments, as well as limitations and possible extensions of the model.

\section{Simulations}
\label{sec:Sim}

\subsection{Computational model}
\label{sec:SimApproach}

\begin{figure}[b]
\centering
\includegraphics[scale=1.0]{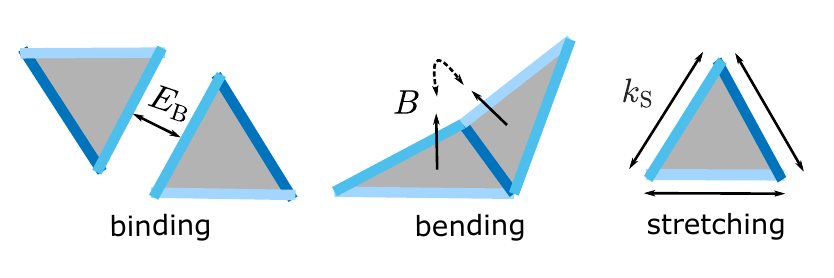}
\caption{ Schematic of the model. The Hamiltonian includes terms that represent edge stretching, monomer-monomer binding, and bending.  Each of the three monomer edges is a different type, and only pairs of edges with the same type can bind. In this work, all edge types have the same binding energy. Each edge type $i$ has a different preferred (`ideal') dihedral angle, $\theta_{\text{id},i}^{(m,n)}$, the set of which determined the target structure $(m,n)$. The energetic cost of deviations from preferred edge lengths and the dihedral angles are controlled by the stretching and bending moduli, $\kStretch$ and $\B$. }
\label{FigSchematic}
\end{figure}

\begin{figure*}[htb!]
\centering
\includegraphics[scale=1.0]{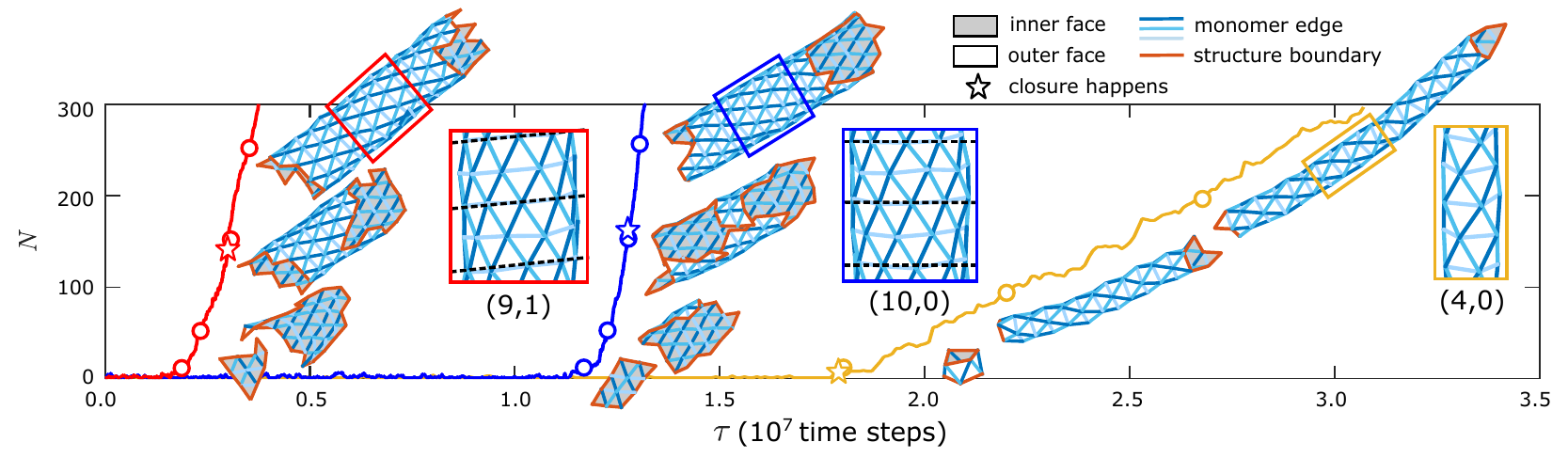}
\caption{Simulation trajectories showing the number of monomers $\N$ in a self-assembled structure as a function of elapsed time $\t$ (number of Monte Carlo sweeps). Snapshots show the configurations at indicated times. Dashed lines in the zoomed-in snapshots label the direction of the edge that is the most perpendicular to longitudinal direction. The assembled tubule geometries are: (9,1) for the red trajectory, (10,0) for the blue trajectory, and (4,0) for the yellow trajectory. The red and blue trajectories have the same target tubule geometry of (10,0) and the same binding energy of $\EB = 6.0\,\kt$. The target tubule geometry for the yellow trajectory is (5,0) and the binding energy is $\EB = 5.0\,\kt$. Other simulation parameters are the same for all the trajectories: $\B=20\,\kt, \fFusion=10^{-3}$.}
\label{FigTraj}
\end{figure*}

In our model, each triangular monomer is composed of three vertices connected by harmonic bonds. The Hamiltonian  is 
\begin{equation}
\label{Hamiltonian}
\begin{split}
H = & \sum_\mathrm{i \in Bound\ Edge\ Pairs} -\EB + \frac{1}{2} B(\theta_{i} - \theta_{0,i})^2 \\
& + \sum_\mathrm{j \in Edges} \frac{1}{2} \kStretch (l_j - l_{0,j})^2 ,
\end{split}    
\end{equation}
with $\EB$ as the monomer-monomer binding energy (set as a positive constant); $\theta_{i}$ and $\theta_{0,i}$ as the instantaneous and preferred dihedral angle between two monomers bound at a common edge $i$, $\B$ as the bending modulus; $l_j$ and $l_{0,j}$  as respectively the instantaneous and stress-free lengths of an edge $j$; and $\kStretch$ as the stretching modulus. The three monomer edges are inequivalent, and setting $\theta_0$ at each of the monomer edges defines the ground state (\emph{target}) tubule geometry (Fig.~\S1). For simplicity, we set $\EB$ and $l_0$ to be identical among all three edges of a monomer, and we consider only a single monomer species. 
Moreover, motivated by the material properties of DNA origami subunits and proteins, we focus here on the limit of thin sheets, in which the bending deformations are much lower in energy than stretching. Therefore, we set $\kStretch= 200\,\kt/l_0^2$ throughout this study so that the monomer edges are nearly fixed in length, and we vary the bending modulus as a control parameter (Fig.~\ref{FigSchematic}). 

We use  Monte Carlo moves to relax the structure, including vertex moves to relax structural degrees of freedom, monomer association and dissociation moves to model assembly and disassembly, as well as moves to model internal rearrangement events such as the splitting and merging of cracks within a structure. 
All the moves guarantee detailed balance  (see SI Section~\rom{8} and \rom{9} for details about the algorithm and the moves). Provided that this set of movements represents the  transitions that are relevant for actual tubule assembly, with approximately correct relative rates for the different moves, the Monte Carlo trajectories can be qualitatively mapped onto the system dynamics. This mapping can be tested by comparing simulation results against experimental observations of tubule assembly kinetics and the structural ensemble of assembled tubules.
The edge fusion and fission moves, which respectively bind two free edges on the structure boundary or split two edges that are already bound, are particularly important for closure/reopening of the tubule structure. We show below that the rate of tubule closure relative to its growth can significantly affect the assembly pathways. Therefore, we define a control parameter -- \textit{edge fusion rate} $\fFusion$, as the ratio between the attempt frequency of edge fusion/fission moves and the unit timescale (which is set to the frequency of vertex moves).

To model assembly from a dilute system of monomers, for which binding between different tubules is negligible, we consider a single assembling structure in each simulation. In addition, we restrict association and dissociation to individual monomers, since assembly of larger oligomers is rare under these conditions. 
To determine well-defined steady-state distributions, we evolve the system in the grand canonical ensemble, in which the assembling structure exchanges monomers with a bath at a fixed chemical potential $\mu$. This situation approximately describes tubule distributions at a point in a reaction with a corresponding free monomer concentration $\rhoFree = \cSS \exp(\mu/\kt)$ with $\cSS$ the standard state concentration. We set  $\mu=-3\,\kt$ throughout this study.

\subsection{Simulation Results}
\label{sec:SimResults}

\textit{Assembly trajectories.} 
The Monte Carlo trajectories exhibit a rich dynamics which proceeds through a series of stages, including nucleation, closure, and growth. Fig.~\ref{FigTraj} shows snapshots from example simulation trajectories at three different parameter sets. During assembly with a large target tubule diameter (red and blue curves), after an initial period of transient assembly and disassembly, the structure surpasses the critical nucleus size (approximately 5 monomers for these conditions) and grows steadily as a curved two-dimensional sheet. Eventually, the boundary edges at opposite sides of the curling sheet begin to touch, and the edges bind. We denote the first such binding event as the point of tubule \emph{closure}. After closure, the tubule geometry is highly stable and the structure undergoes steady growth from both ends.

Even though the red and blue trajectories have the same target structure, they assemble different tubule geometries, denoted by different pairs of integer numbers $(m,n)$ based on the convention from carbon nanotubes~\cite{lee2010}. Representing a tubule as a curled triangular lattice that closes upon itself, the index $m$ gives the number of lattice sites on one turn around the helix, while $n$ gives the number of lattice sites in the orthogonal direction (along the long axis of the tubule) (See Fig.~\S1 for a schematic of the naming convention and SI section \rom{1}A for a detailed description of the notation). In the trajectory with a small target width (yellow line), tubule closure corresponds to the formation of the critical nucleus, after which the tubule undergoes steady growth.

\textit{Tubule geometry distributions.} We performed simulations over a wide range of parameter values to learn how the tubule morphologies arising from dynamical trajectories depend on the relevant physical parameters, such as the bending modulus $\B$, the diameter of the target tubule geometry $D_0$, and the fusion rate $\fFusion$, which influences the closure kinetics. We measured the distribution of tubule geometries at the end of each simulation. Simulations were performed until the length of the structure $L$ grew to approximately three times the tubule circumference, since the geometry distribution is stable by this point (no geometry fluctuations occur beyond this size). We estimated the distributions from 1000 independent trials at each parameter set. 

\begin{figure}[hbt!]
\centering
\includegraphics[scale=1.0]{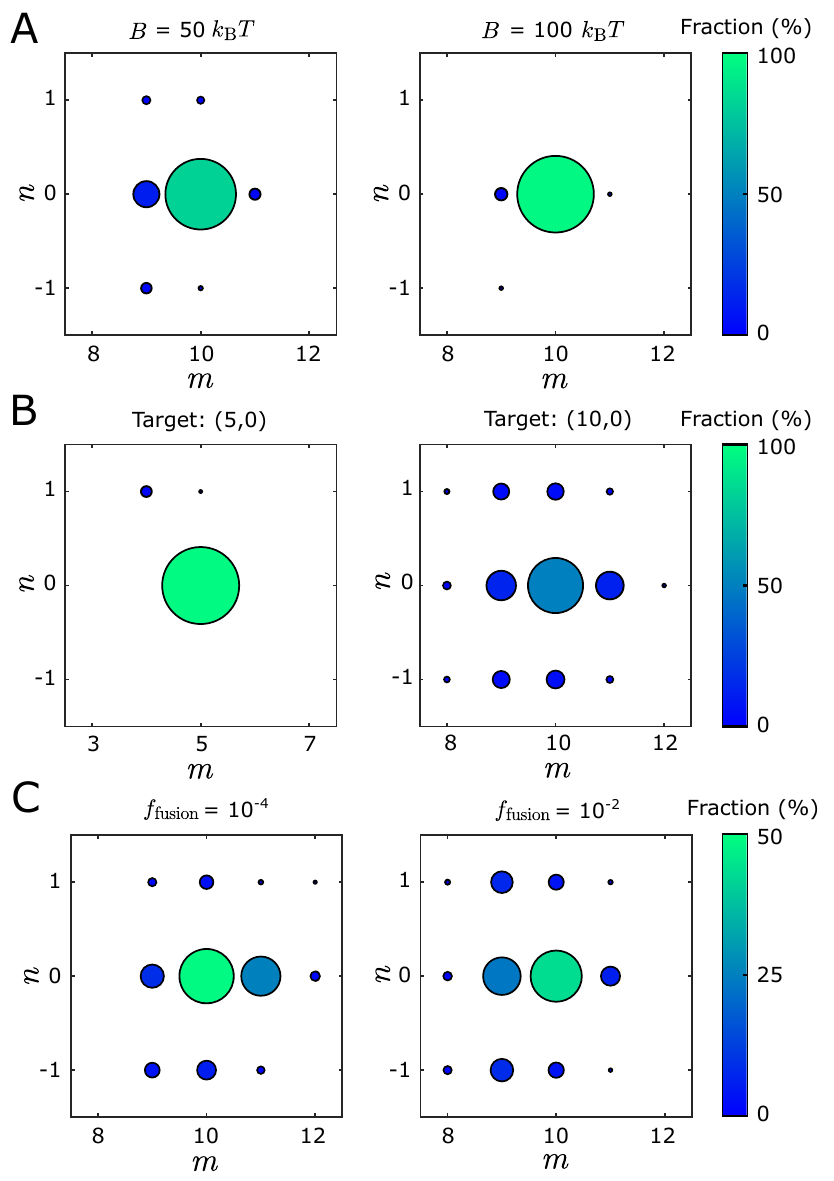}
\caption{Tubule geometry distributions from Monte Carlo assembly trajectories depend on the bending modulus $B$, target tubule geometry, and fusion rate $\fFusion$. The color and size of each circle indicates the fraction of the corresponding tubule geometry within the defect-free population. \textbf{(A)} Geometry distributions for different $\B$ with $\EB= 6\,\kt$, $\fFusion=10^{-3}$, and target tubule geometry (10,0). The fraciton of defect-free tubules is about 97\% for both cases. \textbf{(B)} Geometry distributions for different target geometries with $\B=20\,\kt$, $\EB= 6\,\kt$, and $\fFusion=10^{-3}$. The fraction of defect-free tubules is about 88\% for both cases. \textbf{(C)} Geometry distributions for different $\fFusion$ with $\B=20\,\kt$, $\EB= 6\,\kt$, and target tubule geometry (10,0). As $\fFusion$ decreases from $10^{-2}$ to $10^{-4}$, the fraction of defect-free tubules decreases from 88\% to 52\%, while the fraction of open structures increases from 0 to 45\%. Each distribution in (A)-(C) is estimated from 1000 independent simulation trajectories.}
\label{FigSimWidth}
\end{figure}

We find that tubule structures with different geometries, as well as structures that fail to close, can assemble in the dynamical simulations under the same set of parameter values. We classify the self-assembly outcomes into three categories: \emph{defect-free} tubules, \emph{defective} tubules, and \emph{open structures}. A tubule is \emph{defective} if part of the structure fails to close or multiple tubule geometries are locally identified within the same structure (see Fig.~\S3 and SI section \rom{3}A for identification details). The fraction of defective tubules increases as the bending modulus $B$ decreases and the target diameter $D_0$ increases. To avoid conditions under which defective structures are too prevalent, we set the binding energy to $6\,\kt$ and the monomer chemical potential to $\mu=-3\,\kt$, so that assembly is sufficiently reversible to allow monomer detachment and annealing \cite{Hagan2006,Hagan2014,Fang2020,Whitelam2015}. With these parameters, the fraction of defective tubes is generally below $30\%$. \emph{Open structures} arise when nonuniform curvature causes opposite boundary edges to `miss' the opportunity to bind to each other, leading to a spiral structure that resembles a toilet paper roll (Fig.~\S3). 

The geometry distributions of  defect-free tubules depend on the control parameters. Fig.~\ref{FigSimWidth} shows the distributions of assembled tubules for different bending moduli, target diameters, and fusion rates. The size and color of the circular symbols represent the fraction of different tubule geometries within the defect-free population. Only tubule geometries with populations $\ge 1 \%$ are labeled in the plot. Fig.~\ref{FigSimWidth}(A) shows tubule geometry distributions for two bending moduli $B$, with other parameters fixed. As $B$ increases, the fraction of the target tubule (10,0) increases while the fraction of the off-target tubules decreases. This is consistent with thermodynamics, since the deviations of dihedral angles required for off-target geometries increase in energy with $\B$. Fig.~\ref{FigSimWidth}(B) compares the distributions for two different target geometries. As the diameter of the target geometry $D_0$ increases, the fraction of the target geometry decreases while the fraction and variety of observed off-target geometries increases. This result is also consistent with thermodynamics, since the difference of the ideal dihedral angles between the target state and neighboring tubule geometries is smaller for larger $D_0$ (Fig.~\S2). Therefore, with the same extent of dihedral angle fluctuation, the number of accessible off-target states increases as $D_0$ increases. 
Similar results were described in Ref.~\cite{Videbaek2022}.

Interestingly, even for the same Hamiltonian (in which $B$ and $D_0$ are fixed), changing the edge fusion rate $\fFusion$ changes the skew of the geometry distribution. Fig.~\ref{FigSimWidth}(C) shows that as $\fFusion$ increases from $10^{-4}$ to $10^{-2}$, the geometry distribution changes from skewing above to skewing below the target geometry (10,0). Meanwhile, the proportion of open structures increases from 0 to around $45\%$ as $\fFusion$ decreases from $10^{-2}$ to $10^{-4}$ (Fig.~\S7). This observation reflects the fact that decreasing the closure rate increases the chance that the two edges `miss' each other. As the two edges grow past one another, the size of a curvature fluctuation required to enable closure becomes increasingly unfavorable energetically, and thus more rare. The continued growth of the structure boundary then leads to the spiraling structure described above.

\subsection{Comparison of simulations and experiments}
\label{sec:SimvsExp}

We now compare the results of our dynamical assembly simulations to recent experiments that motivate our work. We find that the morphology distribution of simulated tubules semi-quantitatively agrees with those observed in the experiments \cite{Hayakawa2022}, suggesting that the model incorporates the essential physics of the experiments. 

\begin{figure}[bht!]
\centering
\includegraphics[scale=1.0]{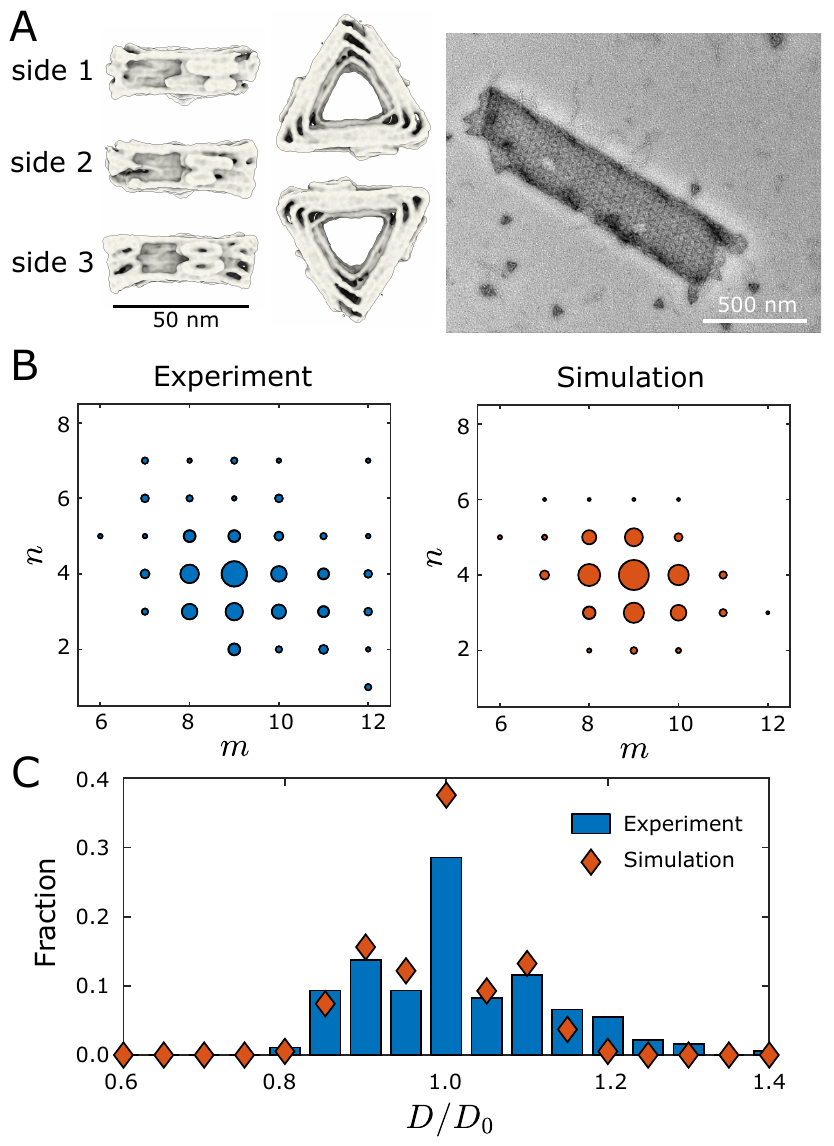}
\caption{Comparing the geometry distributions of tubules assembled in simulations and experiments. \textbf{(A)} Cryogenic electron microscopy reconstructions of a  DNA origami monomer, and a transmission electron microscopy image of an assembled tubule. The left panel shows the monomer under different views. Two monomers bind along their edges through shape-complementary interactions driven by blunt-end DNA base stacking. The right panel shows an assembled (9,4) tubule. Images in \textbf{(A)} were provided by authors of Ref.~\cite{Hayakawa2022}. \textbf{(B)} Tubule geometry distributions measured from experiments~\cite{Hayakawa2022} and simulations. The size of each circle indicates the fraction of the corresponding tubule geometry within the defect-free population. Each simulation data point is estimated from 1000 independent simulation trajectories. \textbf{(C)} Comparing tubule width distributions between experiments (blue bars) and simulations (red symbols). $D_0$ is the diameter of the ideal (9,4) tubule. Simulation parameters: the target geometry is (9,4), $\EB = 6.0\,\kt$, $B=10\,\kt$, and $\fFusion=10^{-3}$.}
\label{FigCompareExpandSim}
\end{figure}

Hayakawa \textit{et al}.~\cite{Hayakawa2022} designed triangular monomers from DNA origami that self-assemble into helical tubules (Fig.~\ref{FigCompareExpandSim}). The monomers interact with each other along their edges through shape-complementary interactions driven by blunt-end DNA base stacking~\cite{Sigl2021}. The interactions are specific --- each monomer edge interacts only with the same edge type on a neighboring subunit.  The bevel angles of the edges of each monomer $\{\thetaIdealTarget\}$ determine the preferred dihedral angles, which, in turn, set the preferred curvatures of the assembly. 

The data set against which we compare our simulation results is obtained from an experimental system that resulted in a most probable tubule geometry of (9,4). 
 Since a key unknown parameter from the experiments is the bending modulus $\B$, we performed simulations with a target geometry of $(9,4)$ at four values of the bending modulus: $B\in\{5, 10, 15, 20\}\,\kt$. All other parameters were fixed to their default values (see section~\ref{sec:Sim}).  We found that assembled structures were highly defective for $\B=5\,\kt$. For $\B\ge10\,\kt$ the majority of tubules were well-formed, with distributions peaked around the target geometry of (9,4). The width of the distribution becomes progressively narrower with increasing $B$, as described in section~\ref{sec:SimResults}. 

We found that a value of $B=10\,\kt$ resulted in a geometry distribution of assembled tubules that closely resembles the distribution observed in the experiments (Fig.~\ref{FigCompareExpandSim}(B)). To facilitate comparison between the two distributions, Fig.~\ref{FigCompareExpandSim}(C) plots the fraction of different tubule geometries against the diameter of the tubules, where  $D_0$ is the diameter of the (9,4) tubule geometry. Although the simulation distribution is slightly narrower than the experiment, we observe that the distributions match fairly closely, especially considering that we have not quantitatively optimized $\B$. Results for the other simulated values of $\B$ are shown in SI Fig.~\S10.


The comparison between simulations and experimental results in Fig.~\ref{FigCompareExpandSim}(B) and SI Fig.~\S10 suggests several important qualitative conclusions: \textit{(1)} the simple model considered here produces results which are semi-quantitatively consistent with those observed in the experiments.  As we will show below, these results can only be explained through a combination of kinetic and thermodynamic effects, which suggests that the highly simplified dynamics of our model captures the most relevant physics. \textit{(2)} It was not possible to directly estimate the bending modulus within the experiments. The simulation results suggest a bending modulus on the order of $10\,\kt$, which is comparable to that of a lipid bilayer membrane ($B/\kt \thicksim 10-20$)~\cite{Otter2003,Dimova2014}. This computational result could be tested in future experiments that measure the distribution of angular fluctuations between monomers. \textit{(3)} Given the qualitative agreement between the computational and experimental results, the simulations can provide a predictive guide for future experiments. We note that a definitive comparison of our simulation results to the experiments, and a precise estimate of the experimental bending modulus, will require additional experimental data sets.  In the subsequent sections, we use the simulations and simple models to understand the effect of relevant control parameters on the morphology distributions of assembled tubules. 

\section{Theoretical models}
\label{Sec:Theory}

To determine whether kinetic effects influence the observed geometry distributions, we compute the equilibrium tubule geometry distribution and compare it against those observed in simulations. We first perform the calculation accounting for the discrete tubule geometries allowed by the finite monomer size, and then we simplify the calculation by adopting the continuum limit. 

\subsection{Discrete equilibrium model}
Motivated by the high rigidity of DNA origami subunits \cite{Sigl2021,Hayakawa2022}, we focus on the regime of high stretching modulus in this work, so that the relative edge length fluctuations are small, $\kt / \kStretch l_0^2 \ll 1$. Thus, in the following calculation we assume that the curvature within a tubule is uniform, and for an assembled geometry $(m,n)$ all the dihedral angles are approximately equal to their ideal value $\thetaIdeal$, with $i\in {1,2,3}$ as the indices of the three sides of a subunit.  
We denote the ideal angles for the target geometry as $\thetaIdealTarget$. The free energy per monomer $g_L^{(m,n)}$ in a tubule geometry $(m,n)$ with length $L$ is then approximately given by
\begin{equation}
\begin{split}
\label{EqZerodE}
   g_{L}^{(m,n)} &= \frac{2\aMono\gamma}{L}-\left(\frac{3\EB}{2}+Ts \right) \\
  & + \frac{1}{4}\B\sum_{i=1,2,3}{(\thetaIdeal-\thetaIdealTarget)^2}    
\end{split}
\end{equation}
in which $\aMono$ is the area of a monomer, $\gamma$ is the line tension accounting for unsatisfied interactions at the two tubule boundaries, $\EB$ is the binding energy, $\B$ is the bending modulus, $T$ is the temperature, and $s$ is the per-monomer entropy. 
The equilibrium probability $P^{(m,n)}$ to assemble the tubule geometry $(m,n)$ with length $L$ is given by
\begin{equation}
P_{L}^{(m,n)} \propto \exp \left[-\beta N(g_{L}^{(m,n)} - \mu)\right],
\label{Discrete}
\end{equation}
where $N$ is the number of monomers in the structure and $\mu$ is the chemical potential. In the grand canonical ensemble $\mu$ is equal to the bath chemical potential, while in the canonical ensemble (conserved total monomer concentration) $\mu = \kt \ln \left(\rhoFree / \cSS\right)$ with $\rhoFree$ the concentration of free monomers.

We consider the large $L$ limit, in which the contribution from the line tension can be ignored, so the free energy per monomer becomes independent of length and will be denoted as $g^{(m,n)}$. Further, at equilibrium, the free energy per monomer of the geometry that minimizes the free energy (in this case the target geometry) is approximately equal to the chemical potential $\mu$~\cite{Hagan2021}\footnote{Note that this approximation is still reasonable in the dynamical simulations, although we set $\mu>\muEq$ so that the simulated tubule can undergo steady-state growth in length $L$, provided that the system is sufficiently close to equilibrium ($\mu\gtrsim\muEq$).}. Since the bending energy of the target geometry is zero, the equilibrium chemical potential is given by $\muEq \lesssim -(\frac{3\EB}{2}+Ts^{*})$, 
where $s^*$ is the entropy per monomer of the target structure. Assuming the entropy is roughly independent of geometry, the probability distribution is then dominated by the bending energy, resulting in
\begin{align}
P^{(m,n)} \propto \exp \left[- \frac{1}{4} \beta N \B \sum_{i=1,2,3}(\thetaIdeal-\thetaIdealTarget)^2\right].
\label{eq:ProbBend}
\end{align}

To test this analysis, we used an adapted thermodynamic integration algorithm to compute the free energy for different tubule geometries $g_{L}^{(m,n)}$. In brief, the algorithm evaluates the free energy change for each geometry along a thermodynamic pathway that gradually transforms the Hamiltonian of the system from a reference state (an Einstein solid with the same number of vertices) to our computation model (Eq.~\ref{Hamiltonian}). We find that the measured free energy difference between different tubule geometries closely agrees with the bending energy difference, confirming the validity of the simplifications described above. See Fig.~\S16 and SI section \rom{7} for details about free energy computations and the comparison to the bending energy.

\subsection{Continuum equilibrium model}
To obtain an approximate analytical expression for the tubule width distribution, we adopt the continuum limit and neglect the presence of defects. In this limit, the bending energy as a function of tubule diameter $D$ is given by the Helfrich energy \cite{helfrich1986}%
\begin{equation}
g^{D} = \frac{2\aMono\gamma}{L} - \left(\frac{3E_\text{B}}{2} + Ts \right) + 2\Tilde{B}\aMono \left( \frac{1}{D} - \frac{1}{D_0} \right)^2,    
\end{equation}
with $D_0$ as the diameter of the target structure and $\Tilde{B}$ as the effective bending modulus in the continuum limit. The continuum bending modulus is related to the bending modulus $B$ of the discrete model by $\Tilde{B} = (\sqrt{3}/2)B$~\cite{Seung1988}.

We evaluate the equilibrium width fluctuations of the closed tubules ($\Delta D \equiv \sqrt{\langle (D - \langle D \rangle )^2\rangle}$) as a function of their length $L$. 
 By performing analogous simplifications to the discrete model  (see SI section \rom{2}), we obtain:
\begin{equation}
\label{eq:Fluc_final}
\Delta D \cong  \sqrt{\frac{\sqrt{3}D_0^{3} \kt}{6\pi B L}} 
\end{equation}.

Equation~\eqref{eq:Fluc_final} shows that the relative equilibrium width fluctuations decrease with bending modulus, but increase with target diameter as $\Delta D/D_0 \thicksim D_0^{1/2}$, as found for spherical curvature-controlled capsids \cite{Hagan2021}. However, a key difference for tubules is that the equilibrium fluctuations become negligible for tubules with large aspect ratios $L \gg D$.  Thus, the observations from simulations and experiments of appreciable width fluctuations in large-aspect ratio tubules indicate that kinetic effects are important in determining the polymorphism.

\subsection{Comparing simulation results against equilibrium width distributions} 
\label{sec:SimVsEquil}

\begin{figure}[hbt!]
\centering
\includegraphics[scale=1.0]{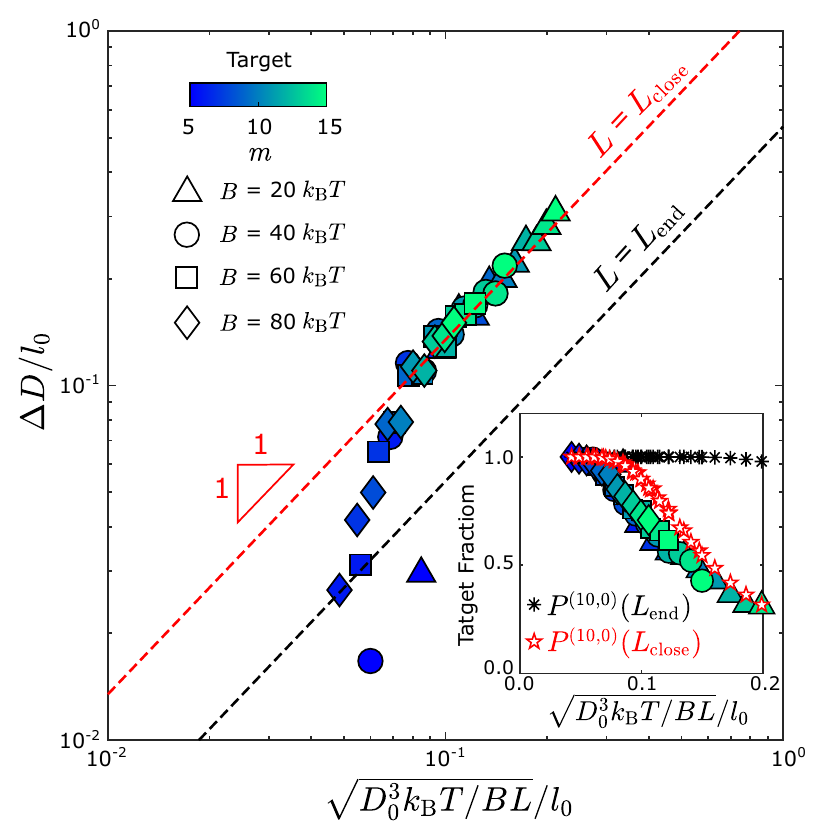}
\caption{Comparing the tubule geometry distribution from Monte Carlo assembly trajectories with the equilibrium theory shows that tubule closure fixes tubule geometries out of equilibrium. 
Tubule width fluctuations $\Delta D$ measured from simulations at different parameter sets, plotted according to the scaling from equilibrium theory (Eq.~\eqref{eq:Fluc_final}).  The dashed black line shows the equilibrium result for the tubule length at the end of the simulation (with  $L=\LEnd$ in Eq.~\eqref{eq:Fluc_final}), while the dashed red line shows the expected result if the geometry is quenched at the point of closure (with $L=\LClosure$ in  Eq.~\eqref{eq:Fluc_final}). Different symbols represent different bending modulus values $\B$, and the color shows the first lattice number $m$ of the target tubule geometry ($m,n$); all structures in this dataset have $n=0$. The inset shows an analogous comparison for the fraction of tubules within the defect-free population that have the target geometry. The black asterisk symbols show the discrete model prediction (Eq.~\eqref{eq:ProbBend} with $L=\LEnd$) and the red pentagon symbols show the discrete model prediction with $L=\LClosure$. Other simulation parameters: $\EB= 6\,\kt$ and $\fFusion =10^{-3}$.
}
\label{FigCompareEqScale}
\end{figure}

By comparing the simulation and equilibrium computation, 
we find that tubule geometry distributions from a dynamical simulations have larger variances than predicted by the equilibrium models. 
Fig.~\ref{FigCompareEqScale} compares the width fluctuations $\Delta D$ measured in the simulations to the scaling law (Eq.~\eqref{eq:Fluc_final}) from the equilibrium computation for different values of the bending modulus and target geometry.  In general, we see that the distributions observed in simulations have larger variances than the equilibrium results. Importantly, the observed  $\Delta D$ collapse to the equilibrium scaling with respect to the target diameter $D_0$ and bending modulus $B$, but  not at  the tubule length at which the geometry measurements are performed ($\LEnd\sim3\pi D_0$, black dashed line in Fig.~\ref{FigCompareEqScale}). The measured diameter fluctuations are much larger than the equilibrium value. Instead, the fluctuations are roughly consistent with the equilibrium prediction for the smaller value of $\LClosure \sim1.5D_0$ at which the tubules closed. Indeed, the results match the equilibrium prediction with $L=\LClosure$ for all parameter values except $\Delta D/l_0 \lesssim 0.5$; below this threshold the fluctuations are smaller than the discrete monomer size and the continuum approximation breaks down. Note that similar results were observed for the same computational model in Ref.~\cite{Videbaek2022} and were shown to be consistent with experiments on DNA origami subunits assembling the tubules in Ref.~\cite{Hayakawa2022}. A detailed description of how we measure the tubule diameter and the closure size is given in the SI (Fig.~\S4 and Fig.~\S5).

The fact that the fluctuations are consistent with the equilibrium prediction, but at the smaller length $\LClosure$, indicates that the geometry distribution is \emph{kinetically} controlled. This conclusion is consistent with the observation from simulations that the geometry rarely changes once a tubule closes, which can be understood from the fact that, after closure, all monomers have their maximum number of bonds except those at the two tubule ends. Rearrangement of the tubule geometry requires breaking a significant number of bonds, and thus overcoming a large free energy barrier.  Note that the substitution of $\LClosure$ into Eq.~\eqref{eq:Fluc_final} amounts to a quasi-equilibrium assumption: Because assembly occurs near equilibrium for the parameters considered in Fig.~\ref{FigCompareEqScale}, the tubule geometry distribution at the time of closure is merely consistent with the equilibrium distribution at the corresponding tubule length $\LClosure$. However, this condition breaks down for larger values of the edge fusion rate $\fFusion$ as discussed next. 

\begin{figure}[hbt!]
\centering
\includegraphics[scale=1.0]{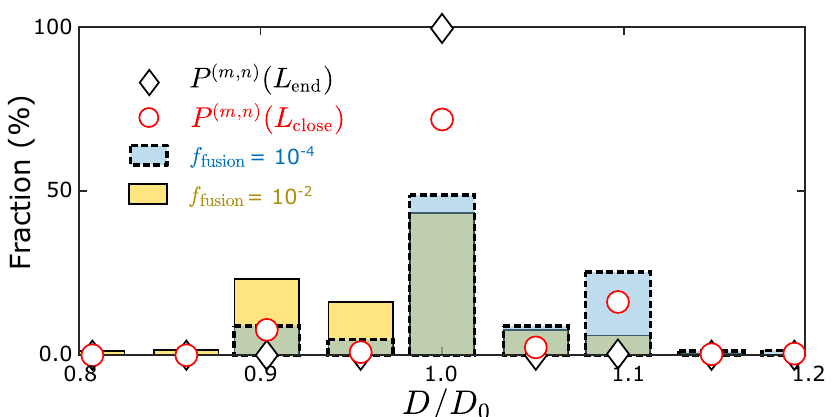}
\caption{Comparing measured tubule width distributions to the discrete equilibrium model (Eq.~\eqref{eq:ProbBend}). Bars are the simulation results with indicated values of $\fFusion$, while symbols represent the equilibrium results, with tubule length at the simulation endpoint or at closure respectively. Other simulation parameters: $\B=20\,\kt$, $\EB=6\,\kt$, and the target tubule geometry is (10,0).
}
\label{FigWidthDis}
\end{figure}

We also compared the fraction of each tubule geometry $(m,n)$ predicted by the discrete model against the simulation results, which indicated a similar trend as for the continuum model: The distribution computed using $\LClosure$ is much closer to the simulation results as compared with using $\LEnd$ (Fig.~\ref{FigWidthDis}), in terms of the width distributon and the yield of the target geometry (Inset of Fig.~\ref{FigCompareEqScale}). Here, we replot the distribution in Fig.~\ref{FigSimWidth}(C) against the diameter of the assembled tubule geometries (bars in Fig.~\ref{FigWidthDis}). However, as $\fFusion$ increases from $10^{-4}$ to $10^{-2}$, the skewness of the tubule width distribution changes from below to above $D_0$. The equilibrium computation at $\LClosure$ does not predict the change in skewness resulting from the change in the assembly kinetics. 

This result shows that the simple picture based on a quasi-equilibrium morphology distribution at $\LClosure$ does not capture all kinetic effects that control the tubule morphology distribution. In section~\ref{sec:KineticModel} we develop a model that accounts for these additional dynamical influences.

\subsection{Kinetic model for tubule geometry distributions}
\label{sec:KineticModel}

The results shown thus far suggest that factors affecting the size and geometry of the tubule at the moment of closure are the key determinants of the observed steady-state geometry distribution. In this section, we develop a discrete model that incorporates both the kinetics and thermodynamics of the system and we show that it semi-quantitatively describes the simulation results. We present an analogous continuum model in SI section \rom{5}C.

We consider the structure before closure as a circular disk that is bent to have the stress-free curvature of the target tubule (see Fig.~\S11 for a schematic of the model). The size $N$ of the disc grows with a rate $\kg$ that is proportional to its boundary length: 
\begin{equation}
    \kg(N) = \kgo \sqrt{N},
\end{equation}
where $\kgo$ is a factor that  depends on $\B$ and $\EB$  (See Fig.\S8 and SI section \rom{3}E for details about growth rate measurements). This estimate is valid when the critical nucleus size is small compared to the closure size, which covers most of the parameter space that we consider in this work. To simplify the model, we ignore the stochasticity in subunit association by assuming that  $N$ increases by one subunit at regular time intervals given by $\Delta t = 1/\kg$.  

While remaining at a size $N$, the open structure also attempts to close with a rate $\kc$. Once the structure closes, we assume that it does not reopen. Allowing for a finite reopening probability is straightforward, but has a negligible effect on the results for the parameters that we focus on because reopening is rare and/or transient.
We assume $\kc$ decreases exponentially with the free energy barrier to closure $\Delta \GClose$, which arises primarily from the bending elastic energy due to the difference in the curvature of the closed structure and the stress-free structure. SI section \rom{3}G presents estimates and measurements of $\Delta \GClose$.

At a given size $N$, the rate $\kc^{(m,n)}$ of closing into a structure $(m,n)$ is then approximated by
\begin{equation}
\label{closurerate}
     \kc^{(m,n)}(N)=\kc^0\exp \left( -\frac{\Delta \GClose^{(m,n)}(N)}{\kt} \right) \IClose^{(m,n)}(N) .
\end{equation}
Here $\kc^0$ is the closure attempt rate (i.e. the rate in the absence of a barrier), and  $\IClose^{(m,n)}$ is a function that indicates whether a particular structure $(m,n)$ is geometrically compatible with closure at size $N$: $\IClose^{(m,n)} =1$ if it is compatible and $\IClose^{(m,n)}=0$ if it is incompatible (see SI section \rom{5}B for details about the determination of $\IClose^{(m,n)}(N)$).
Assuming that shape fluctuations are fast in comparison to the net growth timescale, the net closure rate $\kcNet$ for a disk with size $N$ is then given by a sum over all accessible geometries as
\begin{equation}
     \kcNet(N) = \sum_{(m,n)} \kc^{(m,n)}(N). 
\label{eq:kcNet}
\end{equation}

Finally, we evaluate the closure probability as a function of time. To simplify the calculation, we assume that the structure is larger than the critical nucleus size, and  that closure is a rare event in comparison to growth. In the absence of closure, the time at which a structure first grows to size $N$ is thus $t_N=\sum_{i=1}^{N-1} 1/\kg(i)$, and the probability that such a structure stays open up for an additional time $\delta t < t_{N+1} - t_N$ is 
%
\begin{equation}
 \Pop(t+\delta t,N) = \Pop(t_N,N) \exp \left[-\kcNet(N) \delta t \right].  
\end{equation}
By summing over smaller sizes, we can compute the probability that a structure has stayed open until size $N$ as
\begin{equation}
\Pop(t_N,N) = \prod_{i=1}^{N-1} \exp\left(-\frac{\kcNet(i)}{\kg(i)}\right).   
\end{equation}
The probability for the structure to close at size $N$ is then given by
\begin{equation}
\label{eq:pclosureN}
\begin{split}
\PClose(N) = \Pop(t_N,N)  \left[1 -\exp\left(-\frac{\kcNet(N)}{\kg(N)}\right)\right] 
\end{split}
\end{equation}

The probability to assemble a  geometry $(m,n)$ is then computed by summing over all sizes $N$ that can close to $(m,n)$
\begin{equation}
\label{eq:pclose}
\PClose^{(m,n)} = \sum_{N=1}^{\NMax}  \PClose(N)\frac{\kc^{(m,n)}(N)}{\kcNet(N)},
\end{equation}
where the second term on the right-hand side is the conditional probability for assembling the geometry $(m,n)$, given that the structure closes at size $N$.
Eq.~\eqref{eq:pclose} shows that the ratio of growth to closure rates, which is a kinetic factor, can significantly affect the tubule geometry distribution. Next, we will compare these predictions against the dynamical simulation results from section~\ref{sec:SimResults}.

\subsection{Testing the kinetic model predictions}
\label{sec:ModPrediction}

\begin{figure*}[hbt!]
\centering
\includegraphics[scale=1.0]{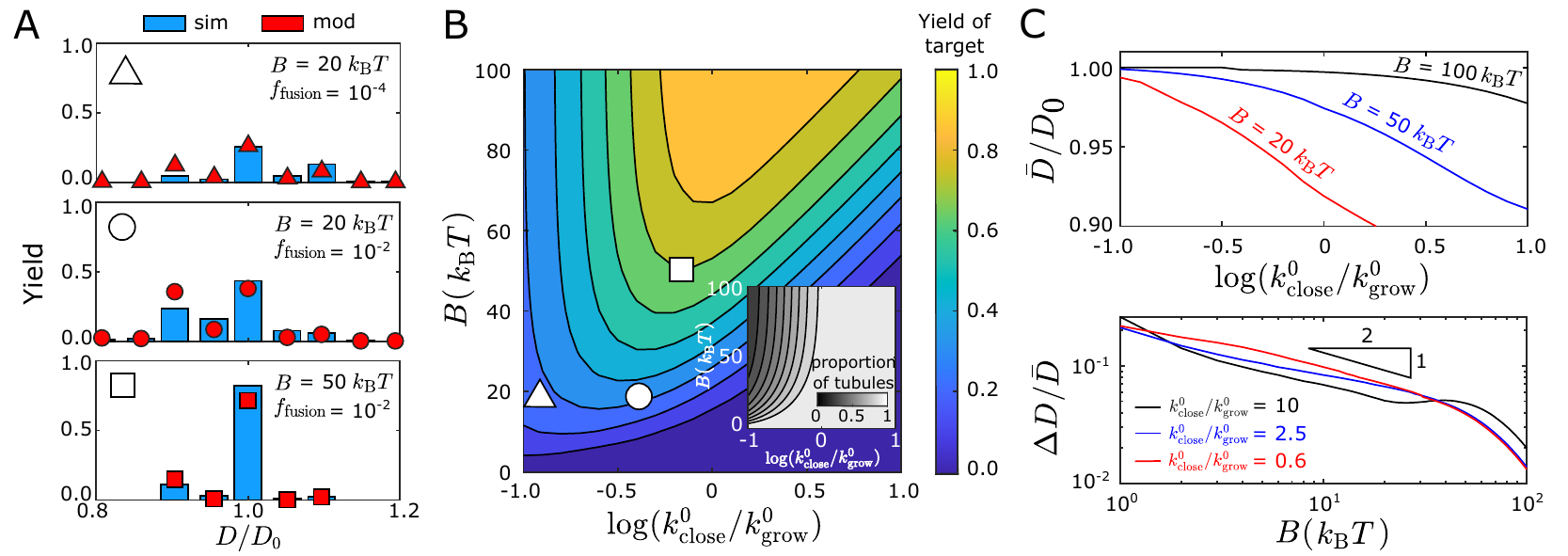}
\caption{The kinetic model captures the geometry distributions observed in simulations.
\textbf{(A)} Comparison of the tubule width distribution between the kinetic model (red symbols) and simulation results (blue bars) for three representative parameter sets. The triangle, circle, or square symbol at the top left of each panel indicates its corresponding location in the parameter space for the plot shown in (B). Yield is defined as the fraction of a tubule geometry assembled within the \textit{entire} population of the assembled structures, including the structures that do not close. \textbf{(B)} Color map showing the yield of the target tubule geometry ($\PClose^{(10,0)}$ defined in Eq.~\ref{eq:pclose}) predicted by the kinetic model as a function of the bending modulus $\B$ and the normalized closure rate $\kco/\kgo$ (shown on a log scale). The yield is taken relative to all assembled structures, including those that fail to close. The inset shows the fraction of closed tubules predicted by the model ($\sum_i^{\infty}\PClose(i)$, $\PClose(i)$ is defined in Eq.~\ref{eq:pclosureN}).\textbf{(C)} Kinetic model predictions for the mean  ($\MeanD$, top panel) and coefficient of variation ($\Delta D/\MeanD$, bottom panel) with respect to the normalized closure rate and bending modulus.  In all cases the target geometry is (10,0). }
\label{Fig5Yield}
\end{figure*}

The simple kinetic model  predicts the tubule geometry distribution and \textit{yield} of the target geometry over a wide range of parameter space of the bending modulus $\B$ and the effective closure rate (normalized by the net growth rate) $\log{(\kc^0/\kg^0)}$. We define the yield  as the fraction of a specific defect-free tubule geometry $(m,n)$ within the entire population (including the unclosed structure and the defective tubules), and $\kc^0$ is defined as the rate for an isotropic open structure to close and form the target geometry.
In the simulations, the effective closure rate $\log{(\kc^0/\kg^0)}$ is controlled by the parameter $\fFusion$, and we measured $\log{(\kc^0/\kg^0)}$ from simulation trajectories as described in SI section \rom{3}F and \rom{3}G. 

The kinetic model accurately predicts the detailed distribution of defect-free tubule geometries, as well as the fraction of structures that fail to close in the dynamical simulations. Comparisons between the kinetic model and the simulation results are shown for three representative parameter sets in Fig.~\ref{Fig5Yield}(A). Starting from the top panel, we reduce $\fFusion$ by $100\times$ at fixed $\B$  (middle panel), which does not significantly change the spread of the distribution of closed tubules, but changes the skew from wider than targeted to narrower than targeted. More significantly, the yield of the target geometry decreases from 40\% to 28\% while the fraction of the target geometry within the defect-free population does not significantly change. This observation is because the proportion of unclosed structures increases from 0 to $\sim45\%$ within the entire population (Fig.~\S7). The kinetic model captures this trend.

In the bottom panel, we increase $\B$ at fixed $\fFusion$ (relative to the middle panel), which narrows the distribution considerably and shifts the mean toward the target diameter. In particular, a significant fraction of tubules with sizes $<D_0$ at $\B=20\,\kt$ shifts to the target geometry with $D_0$ at $\B=50\,\kt$. This trend reflects a combination of thermodynamic and kinetic effects. Increasing $\B$ increases the thermodynamic stability of the target geometry relative to competing structures. It also decreases the net growth rate $\kgo$ because monomer association incurs a greater entropy penalty (since fewer configurations are accessible for binding at higher $\B$), which increases the effective closure rate (see Fig.~\ref{Fig5Yield}(B)) and thus favors smaller structures. However, the thermodynamic effect dominates in this case and shifts the distribution upward toward $D_0$ (See SI Fig.~\S14 for details).

Fig.~\ref{Fig5Yield}(B) shows that the yield of the target geometry also depends on a combination of thermodynamic and kinetic factors. The value of the bending modulus $\B$ sets an upper limit on the yield, while the yield itself changes nonmonotonically with respect to the normalized closure rate at fixed $\B$. These trends reflect the fact that the bending rigidity determines the spread of the distribution, while the closure rate mostly influences the mean of the distribution. As noted above, a higher closure rate leads to structures that close earlier and thus shifts the mean toward smaller structures. 

The bending rigidity controls both the mean $\MeanD$ and the width fluctuation $\Delta D$ of the distribution, while the effective closure rate mostly influences $\MeanD$. Figure.~\ref{Fig5Yield}(C) shows the mean of the distribution $\MeanD$ (top panel) and the width fluctuation $\Delta D$ (bottom panel) as functions of the closure rate and the bending modulus. We see that the mean width monotonically decreases with increasing normalized closure rate or decreasing $\B$. In contrast, the fluctuations $\Delta D/\MeanD$ decrease with $\B$ but depend only weakly on the closure rate. The latter trend is consistent with the qualitative results from the quasi-equilibrium model (section~\ref{sec:SimVsEquil}) based on the equilibrium geometry distribution at the time of closure. In particular, the scaling result  $\Delta D \thicksim \B^{-{1}/{2}}$ still applies. However, the results for the mean width reflect the fact that the probability for an open structure with size $N$ is determined by both the effective closure rate at that size and the time for the structure to remain at size $N$. As shown above, the closure rate increases with $\kco$ and decreases with $\B$, while the time for the structure to remain at a given size decreases with increasing $\kgo$. The irreversible nature of tubule closure plays a key role in this trend, since the smaller structures always have the opportunity to close before larger sizes. Thus, increasing the closure rate or extending the time at a given size will cause the entire distribution to shift toward smaller widths. 

Interestingly, the closure rate does not significantly influence $\Delta D/\MeanD$. Although the kinetic effects discussed above change the tubule closure size $\NClosure$, they do not significantly change the relative prevalence of different tubule geometries at a given $N$. Thus, as long as the shift of the distribution away from $D_0$ is not too large, the density of states around the preferred geometry at size $\NClosure$ remains comparable to that around the target geometry. This distribution is then essentially fixed once closure occurs.

Taken together, the computational and theoretical results considered thus far demonstrate that the outcomes of tubule self-assembly are determined by a combination of thermodynamic and kinetic factors.

\section{Conclusions}
\label{sec:conclusions}

In summary, we have used kinetic Monte Carlo simulations and free energy calculations to understand the dynamical assembly of helical tubules. Our simulations reveal how  assembly pathways and the resulting tubule morphologies depend on control parameters. The geometry distribution of assembled tubules predicted by the simulations semi-quantitatively matches the distribution observed in experiments on tubules assembled from DNA origami monomers~\cite{Hayakawa2022}, suggesting that the model captures the key physics of the experimental system. 
 Further, we show that the simulations provide a useful tool to obtain a first-order estimate of the physical parameters of the experimental system, and can serve as a predictive guide for future experiments.

Comparison of the simulation results with an equilibrium calculation showed that the geometry distribution of assembled tubules depends on a balance between thermodynamic and kinetic effects.  While the observed magnitude of the fluctuations in the tubule width $\Delta D$  match the equilibrium scaling ($\Delta D/D_0 \thicksim \sqrt{ D_0 / \B \LEnd}$) with respect to bending modulus $\B$ and preferred diameter $D_0$, the distribution of assembled tubules is significantly broader and \emph{independent} of length ($\LEnd$).
This behavior can be explained by the fact that the tubule geometry becomes fixed shortly after an assembling proto-tubule closes upon itself.
Closure stabilizes monomer interactions except for those at the two open ends of the tubule, and the topology rearrangements required to significantly change the tubule structure would incur a large free energy barrier.
For this reason, the observed geometry distribution fluctuations tend to scale with the tubule length at closure, as $\LClosure^{-1/2}$. For systems in which closure rates are slow in comparison to growth timescales, the resulting tubule morphology distribution is approximately given by the equilibrium distribution at $\LClosure$. However, for systems with faster closure rates (relative to assembly), additional kinetic effects shift the geometry distribution further out of equilibrium. We developed a simple kinetic model which captures these additional effects.


\textit{Model limitations and outlook.}


While our kinetic model closely reproduces the computational results over a wide range of parameter space, it is limited to regimes in which tubule closure occurs above the critical nucleus size. In particular, the model assumes a positive net growth of the assembling tubule and thus is limited to the forward-biased growth phase that occurs beyond the critical nucleus size. In a future work, we plan to study the nucleation behavior in detail, and how the assembly kinetics and geometry distribution change when closure occurs before nucleation.

In this study we have primarily focused on parameters that lead to well-formed tubules, with a low fraction of defective tubules. However, the simulations provide insights into the factors and mechanisms controlling defect formation. For example, analysis of our simulation trajectories suggests that defective tubules frequently arise when closure happens locally and independently at two or more sites on the boundaries, with geometries that are incompatible with the overall tubule geometry. This mechanism results in a local crack between binding sites, which is unable to anneal unless one of the bound edge-pairs breaks. Further, defective tubules become more probable as the tubule growth rate increases. This observation is consistent with the general principles established from other self-assembly reactions and crystallization (e.g. \cite{Hagan2014,Whitelam2015,Agarwal2014}). When growth rates are sufficiently fast  that monomers that associate with strained interactions cannot anneal before additional subunits assemble, defects become locked into the growing structure. 

While some potential mechanisms of defect formation are disallowed by the simplifications of our model and simulations, these mechanisms can be neglected in the DNA origami experiments that motivate our work. In particular, the algorithm does not allow for binding between multiple partially assembled structures, but these events are negligible under the dilute assembly conditions with a substantial nucleation barrier that tend to lead to  productive assembly \cite{Zlotnick2003,Hagan2014,Whitelam2015}. Similarly, we do not consider
 binding of subunits along non-complementary edges because in the experiments \cite{Hayakawa2022}, monomer-monomer interactions were made highly specific using shape-complementary interactions based on blunt-end DNA base stacking, and there is no evidence of significant binding between non-complementary edges in the experiments.
Note that it would be straightforward to extend the model to eliminate these simplifications to describe other systems for which these mechanisms are not negligible.

We also note that a kinetic Monte Carlo algorithm can only be reliably mapped to real dynamics if the move set accounts for all relevant transitions that occur in a given system, with approximately correct relative rates for each move. In this respect it is encouraging that the simulated tubule geometry distribution compares well with experimental observations. However, further comparison against additional data will be required to stringently test the simulated dynamics, and to refine relative rates. In particular, the simulations described here suggest that the rate at which free edges within an assembled tubule bind to each other is an important parameter controlling the closure rate and defect formation.

With the availability of additional experimental data, some of these unknown coarse-grained parameters could be directly estimated from experiments. At the same time, these measurements would provide estimates of unknown experimental parameter values. For example, by optimizing simulation tubule geometry and width distributions against experiments performed at different parameter values (e.g. target geometry, and monomer concentration), we could estimate the bending rigidity, as well as the closure and growth rates. 
Additional experimental techniques could enable directly estimating some coarse-grained parameters in the model. For example, growth rates could be estimated from dynamic light scattering experiments of tubule assembly, while dimerization rates and free energies for specific monomer-monomer edge interactions could be estimated from static light scattering experiments of subunits which each have only a single edge activated for binding ~\cite{Berne2000,Sigl2021,Hayakawa2022}. Angular fluctuations of dimers measured using atomic force microscopy (AFM) ~\cite{roduit2009,thomas2013} or estimated from electron density in cryo-electron microscopy experiments \cite{Hayakawa2022,Sigl2021,Nakane2018} would provide an independent means of estimating the bending rigidity. 

Through combination with such experimental techniques, our computational and theoretical study could be used to improve the design of existing experimental platforms for tubule assembly. Further, analysis of simulation trajectories for a validated model will provide insights into mechanisms underlying assembly of tubules in these systems, and potentially other related systems with helical geometries such as microtubules \cite{Alushin2014,Nogales2000,Cheng2012,Cheng2013,Cheng2013a,Bollinger2018,Bollinger2019,Stevens2017} and filamentous viruses \cite{Bharat2012,Calder2010,Nakano2021,Selzer2020,Su2018,Xu2020}.  More broadly, the simulation approach and kinetic models are generalizable, and thus could be used to provide similar insights into other assembly geometries with different symmetries and mechanisms of self-limitation.

\section{Conflicts of Interest}
There are no conflicts of interest to declare

\section{Acknowledgements}
We acknowledge support from National Institute Of General Medical Sciences R01GM108021 (HF, BT, MFH), NSF DMR-1710112 (HF, WBR), the Brandeis MRSEC NSF DMR-2011846 (HF, BT, WBR, MFH), the Romanian Ministry of Education and Research, CNCS-UEFISCDI, project no. PN-III-P1-1.1-PD-2019-0236, within PNCDI III (BT), and the Smith Family Foundation (WBR). We also acknowledge computational support from NSF XSEDE computing resources allocation TG-MCB090163 and the Brandeis HPCC which is partially supported by the NSF through DMR-MRSEC 2011846 and OAC-1920147. 

\bibliographystyle{rsc}
\bibliography{TubuleAssembly.bib}


\end{document}